\documentclass[11pt,twoside]{article}
\usepackage{asp2010}

\resetcounters

\begin{document}

\title{Unveiling the structure and kinematics of B[e] stars' disks from FEROS and CRIRES spectra}
\author{M.~F.~Muratore,$^{1,2}$ W.~J.~de~Wit,$^3$ M.~Kraus,$^4$ A.~Aret,$^5$ L.~S.~Cidale,$^{1,2}$ M.~Borges~Fernandes,$^{6}$ R.~D.~Oudmaijer,$^{7}$ H.~E.~Wheelwright$^{8}$
\affil{$^1$Departamento de Espectroscop\'ia Estelar, Facultad de Ciencias Astron\'omicas y Geof\'{i}sicas, Universidad Nacional de La Plata, Paseo del Bosque S/N, B1900FWA, La Plata, Argentina}
\affil{$^2$Instituto de Astrof\'{\i}sica de La Plata, CCT La Plata, CONICET-UNLP, Paseo del Bosque S/N, B1900FWA, La Plata, Argentina}
\affil{$^3$European Southern Observatory, Alonso de Cordova 3107, Vitacura, Santiago, Chile}
\affil{$^4$Astronomick\'y \'ustav, Akademie v\v{e}d \v{C}esk\'e Republiky, Fri\v{c}ova 298, 251\,65 Ond\v{r}ejov, Czech Republic}
\affil{$^5$Tartu Observatory, 61602, T\~oravere, Tartumaa, Estonia}
\affil{$^6$Observat\'orio Nacional, Rua General Jos\'e Cristino 77, 20921-400 S\~ao Cristov\~ao, Rio de Janeiro, Brazil}
\affil{$^7$School of Physics and Astronomy, University of Leeds, Leeds LS2 9JT, UK}
\affil{$^8$Max-Planck-Institut f\"ur Radioastronomie, Auf dem H\"ugel 69, 53121 Bonn, Germany}}

\begin{abstract}
We are investigating the circumstellar material for a sample of B[e] stars using 
high spectral resolution data taken in the optical and near-infrared regions with ESO/FEROS and ESO/CRIRES spectrographs, 
respectively. B[e] stars are surrounded by dense disks of still unknown origin.
While optical emission lines from [O\,{\sc i}] and [Ca\,{\sc ii}] reflect the disk conditions close to the star (few stellar radii), the near-infrared data, 
especially the CO band emission, mirror the characteristics in the molecular part of the disk 
farther away from the star (several AU). Based on our high resolution spectroscopic data, we seek to derive 
the density and temperature structure of the disks, as well as their kinematics. This will 
allow us to obtain a better understanding of their structure, formation history and evolution.
Here we present our preliminary results.
\end{abstract}

\section{Introduction}
B[e] stars are found in different evolutionary phases ranging from 
the pre-main sequence Herbig Ae/B[e] stars to the post-main sequence B[e] 
supergiants and compact planetary nebulae \citep{lamers1998}. They are 
characterized by the presence of permitted and forbidden
emission lines in their optical spectra, and a strong infrared excess emission due to circumstellar dust \citep[e.g.][]{zickgraf1986}. 
Some of these stars also show molecular bands in emission in their infrared spectra \citep{mcgregor1988}.
All this points to the existence of a large amount of circumstellar material, but the ki\-ne\-ma\-tics 
and physical conditions in these extended envelopes (winds and/or disks) are still unknown.

It is important to study the emission of the different components of the circumste\-llar medium, because
 they trace regions at different distances from the star, allowing us to obtain information about the structure and kinematics that helps us characterize the circumstellar environment around these stars. 

\section{Observations}
We obtained high-resolution optical spectra 
for a sample of B[e] stars 
using FEROS (R$\sim$48000), a bench echelle spectrograph mounted at the 2.2-m telescope
of the European Southern Observatory (ESO) in La Silla (Chile).
The wavelength range covered by this instrument is 3600 - 9200\,\AA.
The observations were carried out during two missions: 2005 April 19-21 (Hen\,3-298, CPD-52\,9243), 
 and 2008 December 21-22 (HD\,62623, GG\,Car, CPD-57\,2874). 

For the same stars we also obtained near-infrared high-resolution observations with CRIRES (R$\sim$50000), an echelle spectrograph attached to one of the 8m VLT unit telescopes at the ESO site in Paranal (Chile). The spectra extend approximately from 2.276 to 2.326 $\mu $m, and were obtained on 2009 November 29 and December 2, and 2010 April 6.

In both cases the ESO automatic pipeline reduction was adopted, and the telluric and heliocentric velocity corrections were performed.

\section{Results}

\begin{figure}[!ht]
\plotone{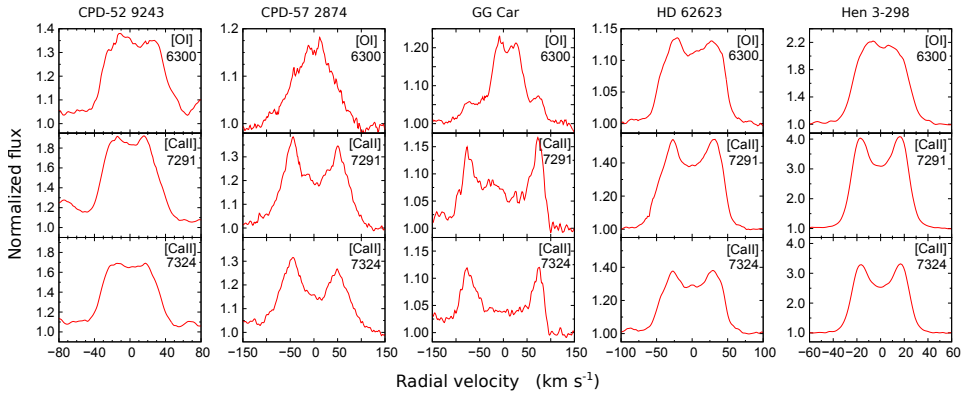}
\caption{Profiles of the [Ca\,{\sc ii}] 7291 \AA, [Ca\,{\sc ii}] 7324 \AA, and [O\,{\sc i}] 6300 \AA \ lines.}\label{fig1}\pagebreak
\end{figure}

The optical spectra display strong emission in the forbidden lines of [O\,{\sc i}] and [Ca\,{\sc ii}]. 
These lines have recently been shown to be valuable tracers of high-density 
disk regions in a sample of B[e] supergiants \citep{aret2012}. Their line profiles are mostly double-peaked, as can be seen in Fig.~\ref{fig1}, indicating either rotation or equatorial outflow.
\\
From their peak-separation (Table \ref{tab:peaksep})
we can extract the kinematics, while a detailed modeling of their line luminosities will provide
information on the density and tem\-pe\-ra\-tu\-re in the line-forming regions \citep[e.g.][]{kraus2007,kraus2010}.
To discriminate which scenario (outflow or rotation) is correct, we
need to know under which conditions the forbidden emission lines are formed.
This has been studied in detail so far for the [O\,{\sc i}] lines \citep{kraus2007,kraus2010}, 
but not much is known about the origin of the [Ca\,{\sc ii}] lines.
Recently, \citet{aret2012} found that for the disks around B[e] 
supergiants the [Ca\,{\sc ii}] lines are formed at distances closer to the star 
(i.e. at higher densities) than the [O\,{\sc i}] lines. If true, the [O\,{\sc i}] line forming 
region should have a lower velocity in the case of Keplerian rotation
and a higher velocity for an outflow scenario. 
Inspection of the velocities listed in Table \ref{tab:peaksep} shows that most stars might be surrounded by a Keplerian rotating disk, while for CPD-52 9243
an outflow scenario might be appropriate. However, detailed modeling of especially the [Ca\,{\sc ii}] lines is necessary before final conclusions can be drawn.
\begin{table}[!ht]
\caption{Peak separations (km s$^{-1}$)}
\vspace{0.3cm}
\label{tab:peaksep}
\smallskip
\begin{center}
{\small
\begin{tabular}{lccc}
\tableline
\noalign{\smallskip}
Object        & [Ca\,{\sc ii}] 7291 \AA & [Ca\,{\sc ii}] 7324 \AA  & [O\,{\sc i}] 6300 \AA \\
\noalign{\smallskip}
\tableline
\noalign{\smallskip}
 CPD-52 9243   &    $30\pm4$  &  $33\pm4$  &   $37\pm4$ \\
 CPD-57 2874   &    $94\pm4$  &  $93\pm2$  &    not resolved  \\
 GG Car        &   $149\pm2$  & $150\pm2$  &   $33\pm2$ \\
 HD 62623      &    $58\pm1$  &  $56\pm1$  &   $50\pm1$  \\
 Hen 3-298     &    $33\pm1$  &  $33\pm1$  &   $14\pm1$  \\
\noalign{\smallskip}
\tableline
\end{tabular}
}
\end{center}
\end{table}

\begin{figure}[!h]
\plotone{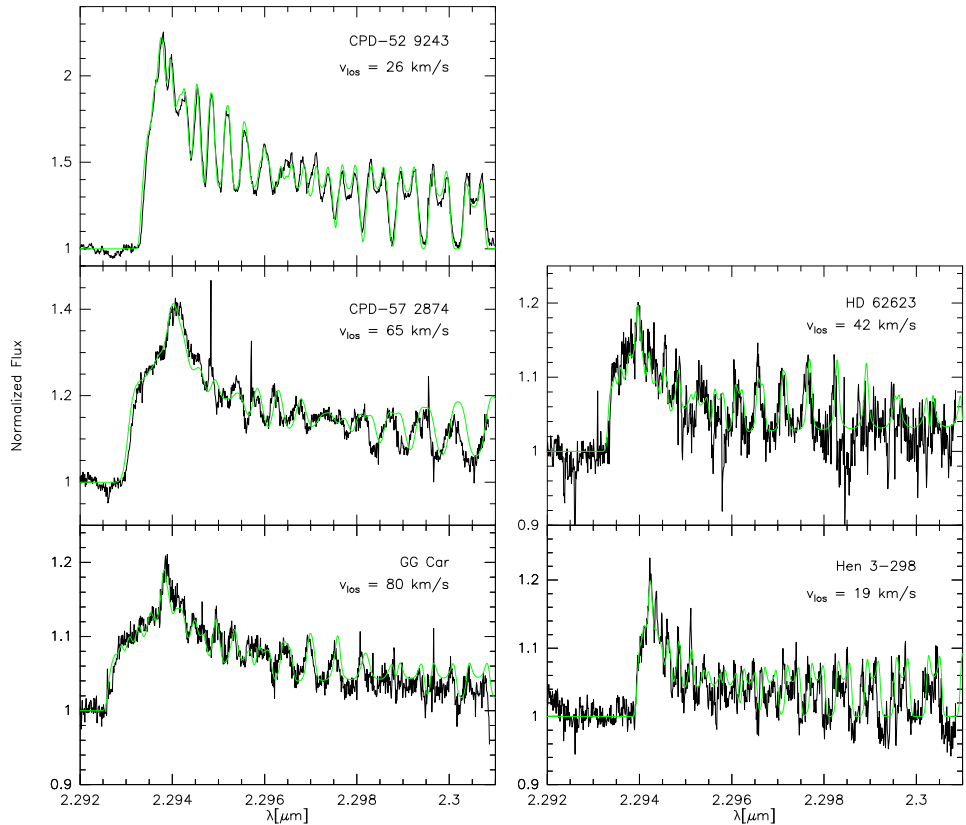}
\caption{Fit (green) to the observed CO bandhead (black). The velocity projected to the  line-of-sight, v$_{\textrm{\scriptsize los}}$, is given in each panel.}\label{fig2}
\end{figure}

The near-infrared spectra display the first bandhead of CO in emission. The shape of the bandhead implies that either rotation or
equatorial outflow broadens the CO band. We applied the model of \citet{kraus2000} to determine the (rotation or outflow)
velocity of the CO gas projected to the line-of-sight. Our preliminary fits are shown in Fig.~\ref{fig2} together with the values obtained for the velocities.

The CO bands typically originate from a very narrow ring representing the inner edge of the molecular disk part \citep[e.g.][]{liermann2010}. Due to its much lower temperature compared to gaseous disk regions, the molecular ring is usually located at (much) larger distances. It is hence surprising that the velocities obtained from the CO bandheads are of similar order or even higher than those of the forbidden emission lines. Whether this discrepancy is real (e.g., caused by the existence of two distinct outflowing rings) or artificial due to the combination of optical and infrared spectra taken at different epochs, needs to be studied in more detail.

\section{Conclusions}
We obtained high-resolution optical and near-infrared spectra of a sample
of B[e] stars. The forbidden emission lines and the CO bandhead show
kinematical broadening due to either rotation or an equatorial outflow.
A detailed analysis of both the optical forbidden lines and the molecular bands will allow us to extract
the structure and the kinematics of the circumstellar material around our stars. This will help to understand the
nature of their disks.

\acknowledgements M.F.M. is a research fellow of the Universidad Nacional de La Plata.
M.F.M. gratefully acknowledges support from an ESO's DGDF 2012 visitor grant.
M.F.M. also acknowledges financial support from the organizers for the participation in this workshop. M.K. acknowledges financial support from GA\,\v{C}R under grant number 209/11/1198.
M.B.F. acknowledges Conselho Nacional de Desenvolvimento Cient\'ifico e
Tecnol\'ogico (CNPq-Brazil) for the post-doctoral grant.

\bibliography{p15-Muratore-ref}

\end{document}